\documentclass[sigconf]{acmart}

\AtBeginDocument{%
  }

\usepackage{multirow}
\usepackage{subcaption}
\usepackage{algorithmic}
\usepackage{listings}

\copyrightyear{2026}
\acmYear{2026}
\setcopyright{cc}
\setcctype{by}
\acmConference[MSR '26]{23rd International Conference on Mining Software Repositories}{April 13--14, 2026}{Rio de Janeiro, Brazil}
\acmBooktitle{23rd International Conference on Mining Software Repositories (MSR '26), April 13--14, 2026, Rio de Janeiro, Brazil}
\acmPrice{}
\acmDOI{10.1145/3793302.3793362}
\acmISBN{979-8-4007-2474-9/2026/04}

\begin{document}

\title[Are Coding Agents Generating Over-Mocked Tests? An Empirical Study]{Are Coding Agents Generating Over-Mocked Tests?\\An Empirical Study}

\author{Andre Hora}
\orcid{0000-0003-4900-1330}
\affiliation{%
  \institution{Department of Computer Science, UFMG}
  \city{Belo Horizonte}
  \country{Brazil}
}
\email{andrehora@dcc.ufmg.br}

\author{Romain Robbes}
\orcid{0000-0003-4569-6868}
\affiliation{%
  \institution{Univ. Bordeaux, CNRS, Bordeaux INP, LaBRI}
  \city{Bordeaux}
  \country{France}}
\email{romain.robbes@labri.fr}

\begin{abstract}
Coding agents have received significant adoption in software development recently.
Unlike traditional LLM-based code completion tools, coding agents work with autonomy (e.g., invoking external tools) and leave visible traces in software repositories, such as authoring commits.
Among their tasks, coding agents may autonomously generate software tests; however, the quality of these tests remains uncertain.
In particular, excessive use of mocking can make tests harder to understand and maintain.
This paper presents the first study to investigate the presence of mocks in agent-generated tests of real-world software systems.
We analyzed over 1.2 million commits made in 2025 in 2,168 TypeScript, JavaScript, and Python repositories, including 48,563 commits by coding agents, 169,361 commits that modify tests, and 44,900 commits that add mocks to tests.
Overall, we find that coding agents are more likely to modify tests and to add mocks to tests than non-coding agents.
We detect that
(1) 60\% of the repositories with agent activity also contain agent test activity;
(2) 23\% of commits made by coding agents add/change test files, compared with 13\% by non-agents;
(3) 68\% of the repositories with agent test activity also contain agent mock activity;
(4) 36\% of commits made by coding agents add mocks to tests, compared with 26\% by non-agents; and
(5) repositories created recently contain a higher proportion of test and mock commits made by agents.
Finally, we conclude by discussing implications for developers and researchers.
We call attention to the fact that tests with mocks may be potentially easier to generate automatically (but less effective at validating real interactions), and the need to include guidance on mocking practices in agent configuration files.
\end{abstract}




\maketitle

\section{Introduction}

Large Language Models (LLMs) have been adopted in multiple software engineering tasks, such as test generation, bug fixing, and refactoring~\cite{fan2023large, monteiro2023end, liang2023can, tufano2023predicting, georgsen2023beyond, hou2023large, hora2024predicting, esem2024_api_migration_llm, di2025deepmig, shirafuji2023refactoring, bouzenia2024repairagent}.
In the context of software testing, numerous studies have investigated the application of LLMs to generate or enhance test cases~\cite{zhang2025large}.
Overall, the results are promising.
For example, LLMs can increase line and branch coverage~\cite{schafer2023empirical, ryan2024code, chen2024chatunitest}, support test-driven development~\cite{mathews2024test, ahmed2024tdd}, enhance existing human-written tests~\cite{alshahwan2024automated}, guide test mutation~\cite{foster2025mutation}, and generate end-to-end tests~\cite{alian2024feature}, to name a few.
Unfortunately, LLM-generated test cases may also present quality issues.
Recent studies have shown that these test cases may exhibit test smells~\cite{ouedraogo2024test, alves2024detecting}, lack assertions, or behave non-deterministically~\cite{alshahwan2024automated} (i.e.,~generating flaky tests~\cite{luo2014empirical}).


In 2025, LLM-based coding agents like Claude Code~\cite{claude}, GitHub Copilot Agent~\cite{copilot}, and Cursor Agent~\cite{cursor} came out and received significant adoption in software development~\cite{li2025rise}.
Unlike traditional LLM-based code completion tools, coding agents work with high \emph{autonomy}; for example, they can invoke external tools, execute code, and complete entire development tasks.
Also, unlike such code completion tools, coding agents leave visible \emph{traces}~\cite{agentminingpaper} in software repositories, such as authoring or co-authoring commits.
Previous work relied on annotations by developers to detect LLM-generated code in software repositories~\cite {tufano2024unveiling, xiao2025self}.
In contrast, agent-based automation becomes observable through the traces left by coding agents~\cite{agentminingpaper}, enabling the analysis of their generated code in the wild, within real-world software systems~\cite{agentadoptionpaper}.

Among their tasks, coding agents may autonomously generate software tests. However, the quality of these tests remains uncertain.
A controversial topic in software testing is the usage of mocking~\cite{meszaros2007xunit}.
Commonly, during testing activities, developers are faced with dependencies (e.g.,~web APIs, databases, file systems, etc.) that make the test harder to implement.
In these cases, developers can use mocking to isolate such dependencies, simulating their behavior, making the test fast, repeatable, and deterministic~\cite{meszaros2007xunit, pereira2020assessing}.
\textbf{Despite the benefits of mocking, its overuse may cause problems: tests can be harder to understand and maintain, and they may provide less assurance that the code is working properly~\cite{google_mock}}.
That is, using mocks in tests only guarantees success if the mocks match the real implementations.
In practice, this is hard to ensure, especially as the real code evolves and possibly gets out of sync with the mocks~\cite{google_mock}.

While the quality of LLM-generated tests has been explored in prior work~\cite{ouedraogo2024test, alves2024detecting, alshahwan2024automated}, the use of mocking remains largely understudied.
As a first step in this direction, an early study evaluated OpenAI's GPT-4o for automating mock decisions by comparing its outputs with developer choices in one system, finding that the LLM often generates more mocks than the developers~\cite{qin2025mock}.
Interestingly, as anecdotal evidence, Kent Beck~\cite{beck2003test} recently stated:
``\emph{[LLM] makes some decisions seemingly at random, like, "Oh, let's use a mock for this test even though the actual object is fine. The price of augmented coding is eternal vigilance}''.\footnote{\url{https://www.linkedin.com/posts/kentbeck_a-problem-im-having-augmented-coding-is-activity-7328422517025452032-Fwxi}}
However, to our knowledge, the agent-generated tests and their relation with mocking have not yet been explored.
Thanks to the traces left by coding agents in software repositories, we are now able to observe this phenomenon~\cite{agentminingpaper}.

This paper presents the first empirical study to investigate the presence of mocks in agent-generated tests of real-world software systems.
We analyzed over 1.2 million commits made in 2025 in 2,168 TypeScript, JavaScript, and Python repositories.
This dataset comprises 48,563 commits by coding agents, 169,361 commits that modify tests, and 44,900 commits that add mocks to tests.
We propose three research questions to address the test generation, mock generation, and mock types:

\begin{itemize}

    \item \textbf{RQ1. How frequently do coding agents generate tests?}
    We find that 60\% of the repositories with agent activity also contain agent test activity. 
    Moreover, 23\% of commits made by coding agents involve the addition or modification of test files, compared with 13\% of commits made by non-agents.
    Considering all commits containing tests, 7\% are made by coding agents, and this proportion increases to 17\% in repositories created in 2025.
    
    \item \textbf{RQ2. How frequently do coding agents generate mocks in tests?} 
    Overall, 68\% of the repositories with agent test activity also contain agent mock activity.
    36\% of test commits made by coding agents add mocks to test code, compared with 26\% of test commits made by non-agents.
    Considering all commits containing mocks, 9\% are made by coding agents, and this proportion increases to 19\% in repositories created in 2025.
    Moreover, in repositories with higher agentic activity, coding agents had a higher ratio of mock commits (36\%) compared to non-agents (28\%).

    \item \textbf{RQ3. What types of mocks are generated by coding agents?}
    Coding agents predominantly use the \emph{mock} type (95\%), whereas non-agents employ a wider variety: \emph{mock} (91\%), \emph{fake} (57\%), and \emph{spy} (51\%).
    In both cases, \emph{dummy} and \emph{stub} are the least adopted types.


\end{itemize}

Overall, we find that coding agents are more likely to modify tests and to add mocks than non-coding agents.
Based on our findings, we discuss several implications for practitioners and researchers.
\emph{First}, the higher likelihood of coding agents modifying test files suggests that these tools are being actively used not only for implementing production code but also for maintaining and expanding tests, highlighting a growing potential for agents to support software testing tasks autonomously.
\emph{Second}, the fact that coding agents tend to rely more heavily on mocking may reflect an overuse of isolation techniques, potentially leading to tests that are easier to generate automatically, but less effective at validating real interactions. 
\emph{Third}, given the tendency of coding agents to rely on mocking, practitioners should explicitly include guidance on mocking best practices and anti-patterns in agent configuration files (e.g., \texttt{CLAUDE.md}) to ensure consistent test generation.
\emph{Lastly}, the limited diversity of test double types used by coding agents suggests that agent-generated tests may rely too heavily on mocks.

\noindent\textbf{Contributions:}
The contributions of this study are twofold:
(1) we provide the first empirical study to analyze agent-generated tests, particularly exploring the generation of tests and mocks in real-world software systems; and 
(2) we propose multiple actionable implications for practitioners and researchers.

\section{Study Design}

\subsection{Case Study}

\subsubsection{Coding Agents}

Coding agents go beyond predictive code completion tools.
Unlike such tools, they can autonomously decompose complex goals, invoke external tools (e.g.,~compilers and search engines), execute code, and refine their outputs with minimal human intervention~\cite{li2025rise, agentadoptionpaper}.
These capabilities mark a shift toward intelligent systems that exhibit decision-making, adaptation, and autonomy in software development~\cite{li2025rise}.
In this paper, we focus on three leading coding agents: Claude Code~\cite{claude}, GitHub Copilot~\cite{copilot}, and Cursor~\cite{cursor}.
These agents are selected based on their growing adoption~\cite{li2025rise, agentadoptionpaper} and their roles as autonomous contributors or co-authors in software repositories.

\subsubsection{Programming Languages}

We aim to perform a multi-language empirical study so we can have a better overview of test and mock adoption in the wild.
We then select the top-3 most used languages on GitHub: Python, JavaScript, and TypeScript~\cite{octoverse-2024}. 


\subsection{Initial Set of Repositories}

Our goal is to analyze real-world, actively maintained repositories hosted on GitHub.
To this end, we rely on the SEART GitHub Search Engine (seart-ghs), a tool that allows researchers to sample repositories to use for empirical studies by using multiple combinations of selection criteria~\cite{Dabic:msr2021data}.
This tool maintains metadata for all GitHub repositories with at least ten stars.
Based on this tool, we selected repositories that meet the following criteria: at least 100 commits, a minimum of 5,000 non-blank lines of code, not being forks, and having at least one commit in the last two months.
This process yielded an initial set of 114,098 repositories.

\subsection{Detecting Repositories with Agents}

From the initial set of repositories, we selected the ones where the main  language was Python, TypeScript, or JavaScript and adopted the coding agents Claude, Copilot, or Cursor.
To detect the presence of such coding agents, we cloned the repositories and verified the existence of files or directories associated with them.
We relied on the patterns presented in Table~\ref{tab:agents-patterns}, derived from the official documentations of Claude~\cite{claude-best-practices}, Copilot~\cite{copilot-best-practices}, and Cursor~\cite{cursor-best-practices}.

\begin{table}[h]
    \centering
    \footnotesize
    \caption{Coding agent file and directory patterns.}
    \begin{tabular}{lp{6cm}}
        \toprule
        \textbf{Coding Agent} & \textbf{File and Directory Pattern} \\ \midrule
        Claude        & \texttt{CLAUDE.md}, \texttt{.claudeignore}, \texttt{.claude/}, \texttt{anthropic/}  \\ \midrule

        Cursor        & \texttt{CURSOR.md}, \texttt{.cursor/}, \texttt{.cursorrules}  \\ \midrule
        
        Copilot     & \texttt{copilot\_instructions.md}, \texttt{copilot-instructions.md}, \texttt{.copilot-*.md}, \texttt{.copilotignore}, \texttt{.copilot/}  \\
        
        \bottomrule
    \end{tabular}
    \label{tab:agents-patterns}
\end{table}

These configuration files are used to document commands, instructions, and guidelines for coding agents.
For example, the Claude Code documentation states~\cite{claude-best-practices}: ``\emph{\texttt{CLAUDE.md} is a special file that Claude automatically pulls into context when starting a conversation. This makes it an ideal place for documenting: common bash commands, core files and utility functions, code style guidelines, testing instructions [...]}''.
We found a total of 2,168 repositories with agent files or directories across the three target programming languages.

\subsection{Detecting Agent Commits}
\label{sec:detecting-agents}

\begin{figure}[h]
     \centering
     \begin{subfigure}[b]{0.45\textwidth}
         \centering
         \includegraphics[width=\textwidth]{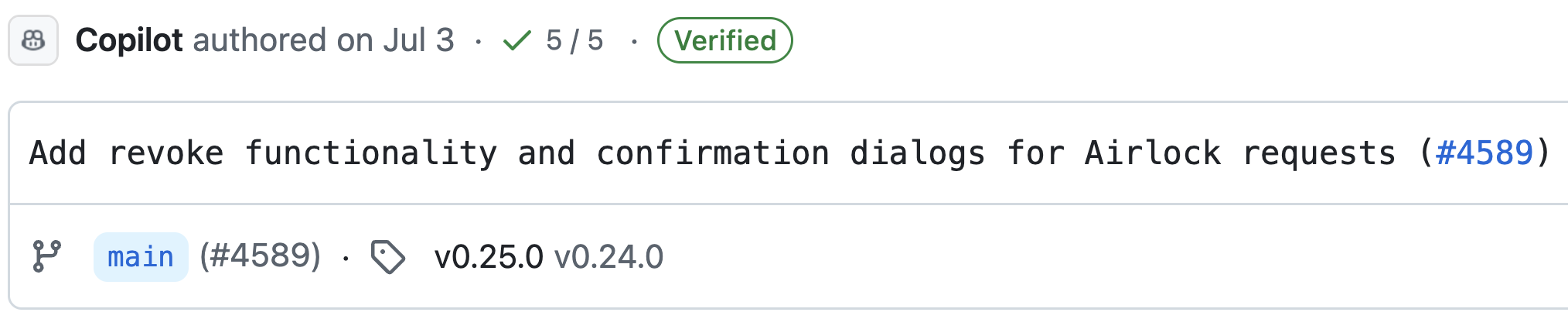}
         \caption{Commit authored by Copilot (microsoft/azuretre).}
         \label{fig:ex1}
     \end{subfigure}
     \par\medskip
     \begin{subfigure}[b]{0.45\textwidth}
         \centering
         \includegraphics[width=\textwidth]{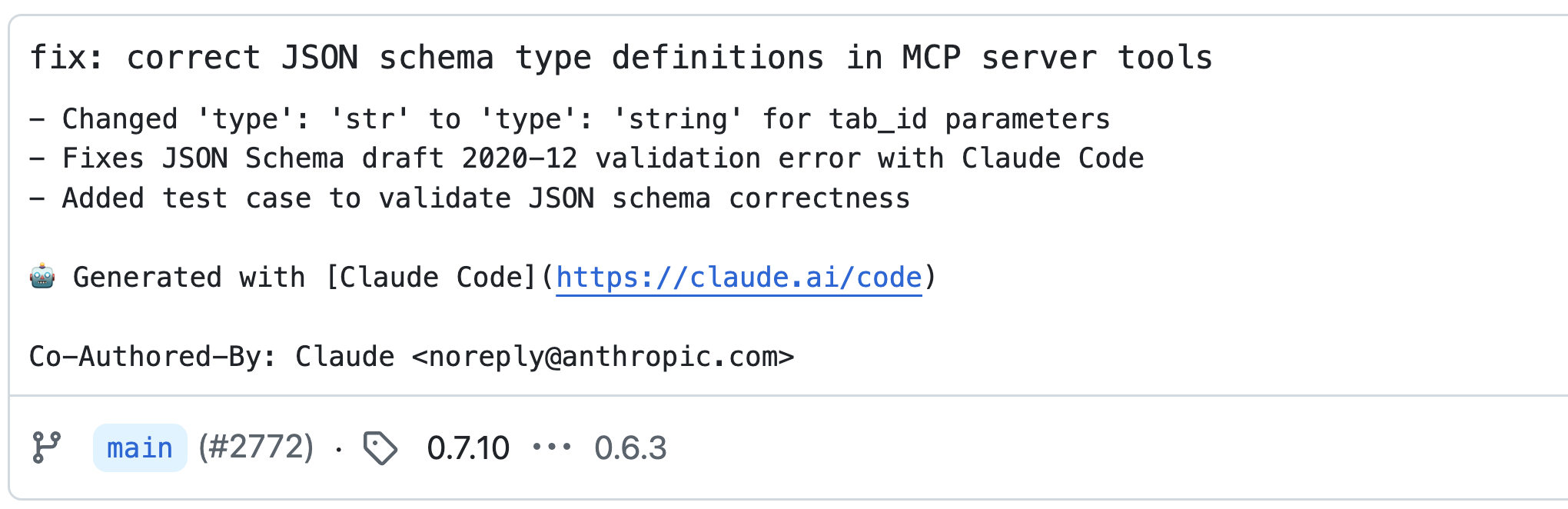}
         \caption{Commit co-authored by Claude (browser-use/browser-use).}
         \label{fig:ex2}
     \end{subfigure}
    \caption{Examples of agent commits.}
    \Description{Examples of agent commits.}
    \label{fig:examples}
\end{figure}

The presence of a coding agent file or directory in a repository does not necessarily imply that the agent was used to make commits.
Therefore, the next step in our study design is to identify repositories containing commits actually made by coding agents.
To this end, we analyzed the commits of the selected repositories, searching for commits authored or co-authored by the coding agents Claude, Cursor, or Copilot.
The \emph{co-authored} metadata is a GitHub feature that allows a commit to have multiple authors~\cite{co-authored-by}.
In this case, developers can attribute a commit to more than one author by adding one or more \texttt{Co-authored-by:} trailers to the commit message, thereby making those co-authors visible on GitHub.
Although GitHub recommends using the \texttt{Co-authored-by:} commit trailer~\cite{co-authored-by}, we observed that coding agents may also use variants such as \texttt{Co-Authored-By:} (see Figure~\ref{fig:ex2}); therefore, we searched for these trailers in commit messages in a case-insensitive manner.

We then identified commits whose \emph{author} or \emph{co-authored-by} metadata contained, in a case-insensitive manner, the substrings \emph{claude}, \emph{cursor}, or \emph{copilot}.
In addition, we also looked for traces of all popular coding agents in commit authors or co-authors, including Aider, OpenHands, Devin AI, Google Jules, Cline, Junie, Gemini, CodeRabbit, and Windsurf.
All patterns used to detect the presence of agents were derived from official documentation or from a manual assessment of commits authored by coding agents.
We manually inspected 500 agent commits and found a precision of 100\% (500 out of 500) in the agent commit classification.

We label commits containing these coding agents as \emph{agent commits}, while other commits as \emph{non-agent commits}.
Figure~\ref{fig:ex1} presents an example of a commit authored by Copilot in the repository microsoft/azuretre,\footnote{\url{https://github.com/microsoft/azuretre/commit/a266b50733}} while Figure~\ref{fig:ex2} shows a commit co-authored by Claude in the repository browser-use/browser-use.\footnote{\url{https://github.com/browser-use/browser-use/commit/b2da2fec89}}

We found a total of 48,563 agent commits in 1,219 repositories.

\subsection{Detecting Test Commits}
\label{sec:detecting-tests}

The next step in our design is to detect \emph{test commits}, that is, commits that add or modify test files.
Table~\ref{tab:test-patterns} presents the patterns used to detect the presence of the test files in Python, TypeScript, and JavaScript.
For Python, we followed the convention that test files begin with \texttt{test\_} or end with \texttt{\_test}, which is used by the testing frameworks Pytest and unittest to automatically detect tests.
For TypeScript and JavaScript, test files typically have the suffixes \texttt{test.ts} or \texttt{spec.ts} (in TypeScript) and \texttt{test.js} or \texttt{spec.js} (in JavaScript), which are used by popular testing frameworks such as Jest and Mocha to automatically detect tests.\footnote{\url{https://create-react-app.dev/docs/running-tests/\#filename-conventions}}
In addition, we also consider as test files all files located in directories whose paths contain the substring \texttt{test}, such as \texttt{test/}, \texttt{e2e\_tests/}, and \texttt{\_\_tests\_\_/}.
For TypeScript and JavaScript, we also include directories that end with \texttt{spec/}.
These directories are commonly used in Python, TypeScript, and JavaScript projects and, depending on the testing framework, may not require files within them to include the \texttt{test} prefixes or suffixes.
Based on this method, we found a total of 169,361 test commits in 1,779 repositories.

\begin{table}[h]
    \centering
    \footnotesize
    \caption{Test file patterns.}
    \begin{tabular}{lll}
        \toprule
        \textbf{Prog. Language} & \textbf{Test File Pattern} & \textbf{Example} \\ \midrule
        Python & \texttt{test\_*.py} or \texttt{*\_test.py} & \texttt{test\_foo.py} \\
        TypeScript & \texttt{*test.ts} or \texttt{*spec.ts} & \texttt{foo.test.ts} \\
        JavaScript & \texttt{*test.js} or \texttt{*spec.js} & \texttt{foo.test.js}\\
        \bottomrule
    \end{tabular}
    \label{tab:test-patterns}
\end{table}

\subsection{Detecting Mock Commits}
\label{sec:detecting-mocks}

According to Meszaros~\cite{meszaros2007xunit}, mocks ``\emph{replace an object the system under test (SUT) depends on with a test-specific object that verifies it is being used correctly by the SUT}''.
In fact, a mock is a particular type of test double~\cite{meszaros2007xunit}.
The testing literature proposes five test doubles: dummy, stub, spy, mock, and fake, each one with its isolation characteristics~\cite{meszaros2007xunit}.
Dummy is the simplest test double, as its behavior is not predefined nor verified;
stub offers predefined implementations;
mock and spy verify interactions; and
fake provides simplified implementations~\cite {fazzini2022use}.
Overall, a test double can be defined as ``\emph{the broadest term available to describe any fake thing introduced in place of a real thing for the purpose of writing an automated test}''.\footnote{\url{https://github.com/testdouble/contributing-tests/wiki/Test-Double}}
Meszaros recognizes that terminology around the types of test doubles is confusing and inconsistent, hence different people use distinct terms to mean the same thing~\cite{meszaros2007xunit}.

Developers tend to use the terms mocks and test doubles interchangeably~\cite{pereira2020assessing}.
In this context, Robert Martin~\cite{martin2009clean} mentions: ``\emph{The word mock is sometimes used in an informal way to refer to the whole family of objects that are used in tests}''.\footnote{\url{https://blog.cleancoder.com/uncle-bob/2014/05/14/TheLittleMocker.html}}

Thus, in this paper, we use the term mock due to its popularity, but our analysis encompasses the full family of test doubles, that is, dummy, stub, spy, mock, and fake.

\subsubsection{Mock Usage}

There are two approaches to using mocks in test code: (1) employing mocking frameworks or (2) manually creating them in the source code~\cite{pereira2020assessing}.

JavaScript and TypeScript provide multiple mocking frameworks such as Sinon.JS,\footnote{\url{https://sinonjs.org}} Jest Mock Functions,\footnote{\url{https://jestjs.io/docs/mock-functions}} and Jasmine Spy.\footnote{\url{https://jasmine.github.io/tutorials/spying_on_properties}}
Examples of APIs provided by such mocking frameworks include \texttt{mock}, \texttt{spyOn}, \texttt{spyOnProperty}, and \texttt{fake}, to name a few.
In Python, the unittest testing framework provides unittest.\-mock,\footnote{\url{https://docs.python.org/3/library/unittest.mock.html}} while Pytest provides monkeypatch\footnote{\url{https://docs.pytest.org/en/stable/how-to/monkeypatch.html}} and pytest-mock.\footnote{\url{https://pytest-mock.readthedocs.io/en/latest/index.html}}
As expected, such frameworks include multiple APIs to create mocks, such as \texttt{MagicMock}, \texttt{mocker}, \texttt{mock}, \texttt{spy}, and \texttt{stub}.
In addition, Pytest provides the \texttt{monkeypatch}\footnote{\url{https://docs.pytest.org/en/stable/reference/reference.html\#pytest.MonkeyPatch}} fixture for mocking functionality in tests, while unittest provides \texttt{patch},\footnote{\url{https://docs.python.org/3/library/unittest.mock.html\#the-patchers}} which acts as a function decorator, class decorator, or a context manager.

In the second approach, developers manually create mocks in test code~\cite{percival2014test}, including classes, methods, and variables with mocked implementations.
For example, a search in the microsoft/vscode repository for the five test double types returns thousands of occurrences, such as \texttt{dummy\-Display\-Obj}, \texttt{Remote\-Service\-Stub}, \texttt{spy\-Encryption\-Service}, \texttt{Mock\-Label\-Service}, and \texttt{\_fake\-Auth\-Populate}.\footnote{GitHub search: \url{https://github.com/search?q=repo\%3Amicrosoft\%2Fvscode+language\%3ATypeScript+dummy+OR+stub+OR+spy+OR+mock+OR+fake&type=code}}

Thus, our approach to automatically detecting mock usage aims to capture both framework-based and manually created mocks.
Specifically, to determine whether a block of code uses a test double, we first parse the code to extract all identifiers and then check whether any of them include the terms \emph{dummy}, \emph{dummies}, \emph{stub}, \emph{mock}, \emph{spy}, \emph{spies}, or \emph{fake}, in a case-insensitive manner.
In addition, we also verify the presence of the terms \emph{monkeypatch} and \emph{monkey\_patch}.
Moreover, since \emph{patch} is a common term that may appear in contexts unrelated to test doubles, we only capture its occurrences when used in Python decorators (i.e.,~\texttt{@patch}, \texttt{@mock.patch}, or \texttt{@unittest.mock.patch}) or Python context managers (i.e.,~\texttt{with patch}, \texttt{with mock.patch}, or \texttt{with unittest.mock.patch}), which are standard ways to invoke the \texttt{patch} API according to the official unittest documentation.
Identifiers containing such mock-related terms are referred to as \emph{mock identifiers}, e.g.,~\texttt{dummy\-Display\-Obj}.

We manually inspected 100 randomly selected mock commits of 10 repositories (10 commits per repository), and found a precision of 94\% (94 in 100) in the identifier-based mock detection.

This approach is in line with prior studies, which also rely on code identifiers to detect the presence of test doubles~\cite{pereira2020assessing, fazzini2022use}.
Therefore, we can capture the presence of test doubles in any code constructor, including classes, methods, functions, parameters, arguments, attributes, calls, and variables.

\subsubsection{Mock Commits}

The next step in our study is to identify commits that actually add mocks to test code; for brevity, we refer to these as \emph{mock commits}.
It is well known that commits often include unrelated modifications—also referred to as tangled changes~\cite{dias2015untangling}. For instance, a single commit may simultaneously add a new test and refactor an existing one. To mitigate potential noise introduced by such tangled commits, we focus on identifying mock commits that explicitly introduce new mock identifiers.

Specifically, for each test commit, we examine the test files that were added or modified.
First, we extract all mock identifiers appearing in the added and deleted lines of these files, yielding two sets: \emph{added mock identifiers} and \emph{deleted mock identifiers}.
Next, we determine the \emph{new mock identifiers}, defined as the mock identifiers that appear only in the added lines (i.e., not present among the deleted ones).
Finally, we classify a test commit as a \emph{mock commit} if it introduces at least one \emph{new mock identifier}.
Based on this method, we found 44,900 mock commits in 1,381 repositories.

\begin{figure}[h]
     \centering
     \begin{subfigure}[b]{0.47\textwidth}
         \centering
         \includegraphics[width=\textwidth]{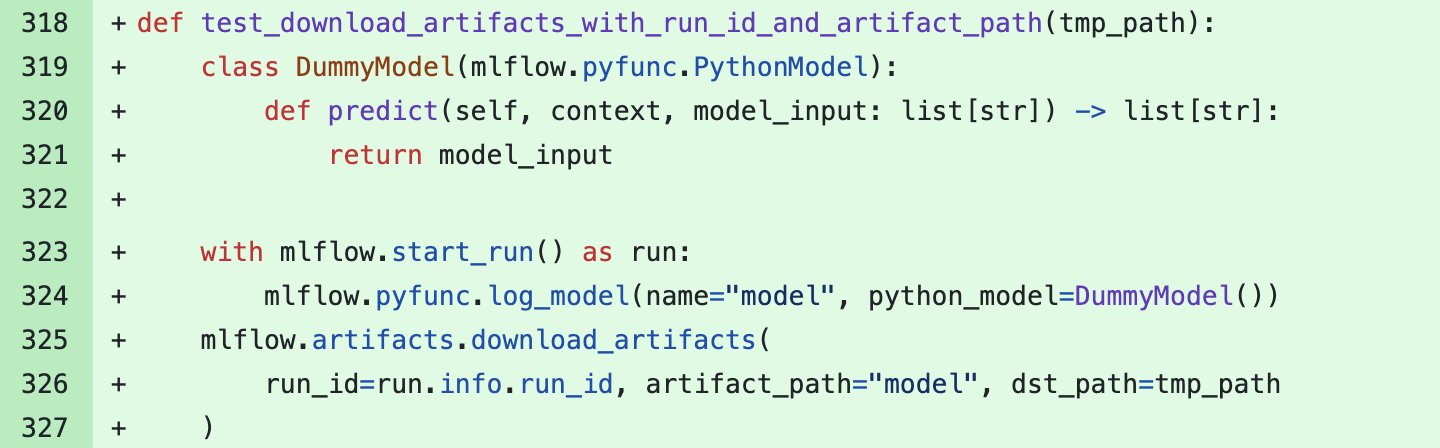}
         \caption{New mock identifier: \texttt{DummyModel} (mlflow/mlflow, 6ba3cefcee).}
         \label{fig:mock-commit-ex1}
     \end{subfigure}
     \par\medskip
     \begin{subfigure}[b]{0.47\textwidth}
         \centering
         \includegraphics[width=\textwidth]{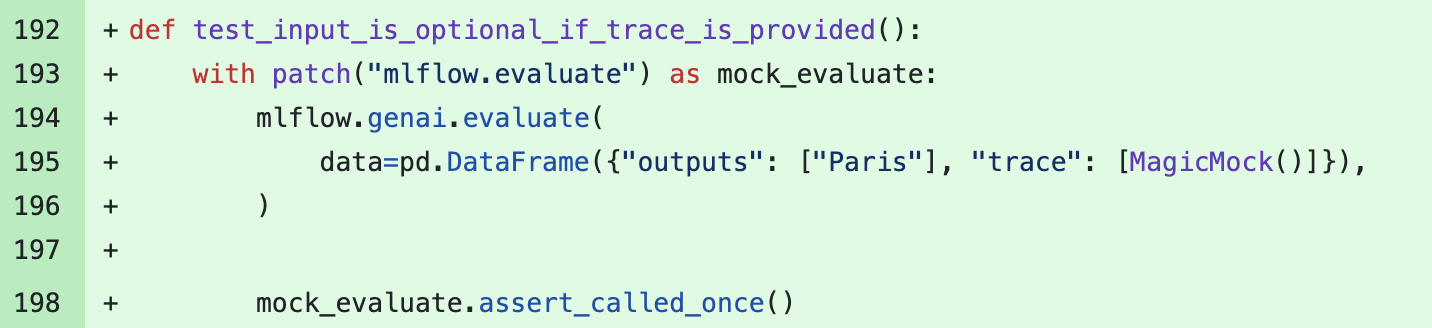}
         \caption{New mock identifiers: \texttt{mock\_evaluate}, \texttt{MagicMock}, and \texttt{patch} (mlflow/mlflow, 6920a3a779).}
         \label{fig:mock-commit-ex2}
     \end{subfigure}
     \par\medskip
     \begin{subfigure}[b]{0.47\textwidth}
         \centering
         \includegraphics[width=\textwidth]{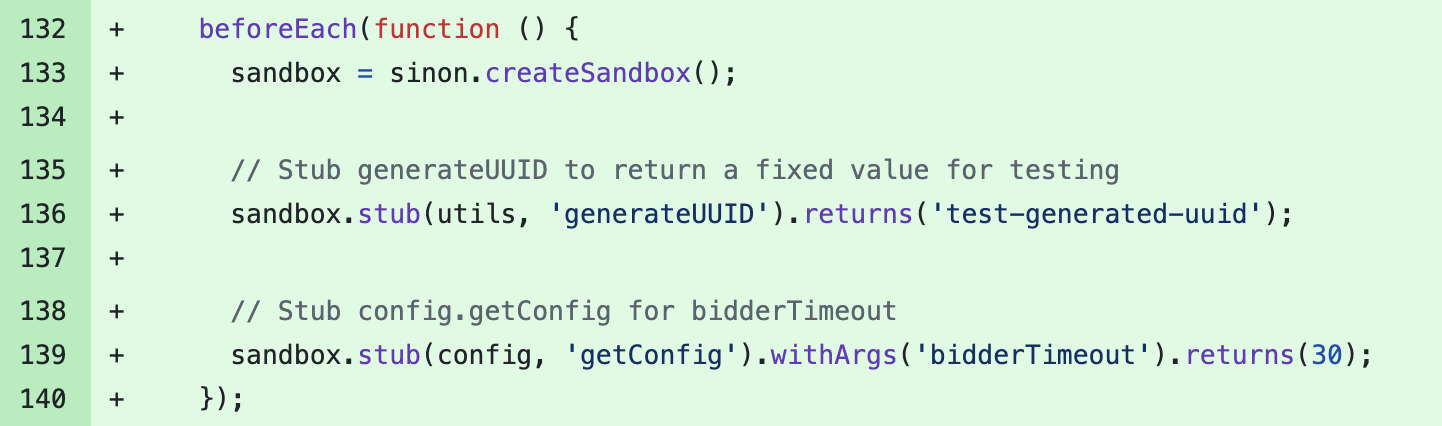}
         \caption{New mock identifier: \texttt{stub} (prebid/prebid.js, 5e57caa8ce).}
         \label{fig:mock-commit-ex3}
     \end{subfigure}
    \caption{Examples of mock commits.}
    \Description{Examples of mock commits.}
    \label{fig:mock-commits}
\end{figure}

Figure~\ref{fig:mock-commits} presents three examples of mock commits.
First, we see a mock commit in the repository mlflow/mlflow that added the manually created test double \texttt{DummyModel} (Figure~\ref{fig:mock-commit-ex1}).\footnote{Mock commit: \url{https://github.com/mlflow/mlflow/commit/6ba3cefcee}}
Next, in Figure~\ref{fig:mock-commit-ex2}, we observe a mock commit that added two mock APIs (\texttt{MagicMock} and \texttt{patch} of unittest.mock) and the alias \texttt{mock\_evaluate}.\footnote{Mock commit: \url{https://github.com/mlflow/mlflow/commit/6920a3a779}}
Lastly, in the repository prebid/prebid.js we see a mock commit that added the mock API \texttt{stub} of the Sinon.JS mocking framework (Figure~\ref{fig:mock-commit-ex3}).\footnote{Mock commit: \url{https://github.com/prebid/prebid.js/commit/5e57caa8ce}}

\subsection{Dataset Overview}

Table~\ref{tab:summary-repos-commits} summarizes the analyzed data.
Overall, we analyzed 1,254,878 commits made in 2025 in 2,168 repositories.
Among these, 48,563 were agent commits from 1,219 repositories, 169,361 were test commits from 1,779 repositories, and 44,900 were mock commits from 1,381 repositories.
To perform the repository analysis, we adopted PyDriller~\cite{pytest} and GitEvo~\cite{gitevo}.
Our scripts and dataset are publicly available at: \url{https://doi.org/10.5281/zenodo.17427638}.

\begin{table}[h]
    \centering
    \footnotesize
    \caption{Summary of commits and repositories.}
    \begin{tabular}{lrrrr}
        \toprule
        \textbf{Commits} & \textbf{TS} & \textbf{JS} & \textbf{Python} & \textbf{Total} \\ 
        \midrule
        All commits & 835,781 & 98,389 & 320,708 & 1,254,878 \\
        Agent commits & 32,728 & 3,892 & 11,943 & 48,563 \\
        Test commits & 94,747 & 6,428 & 68,186 & 169,361  \\
        Mock commits & 23,838 & 1,561 & 19,501 & 44,900 \\
        \midrule
        \multicolumn{5}{l}{\textbf{Repositories}} \\ 
        \midrule
        With agent files or directories & 1,392 & 242 & 534 & 2,168 \\
        With agent commits & 773 & 117 & 329 & 1,219 \\
        With test commits & 1,149 & 143 & 487 & 1,779 \\
        With mock commits & 890 & 89 & 402 & 1,381 \\
        \bottomrule
    \end{tabular}
    \label{tab:summary-repos-commits}
\end{table}



\subsection{Research Question}

\subsubsection{RQ1. How frequently do coding agents generate tests?}

First, we investigate how frequently coding agents add or modify test files.
We examine the overlap between agent/non-agent commits and test/non-test commits in all 1,254,878 commits.
To assess whether these two categorical variables are statistically associated, we applied a Chi-squared test of independence.
\textbf{Rationale:} This analysis is a first step toward better understanding the extent to which coding agents contribute to testing activities compared to non-agents.

\subsubsection{RQ2. How frequently do coding agents generate mocks in tests?}

To better understand how coding agents generate mocks, we divide this research question into two parts: (1) at the commit level and (2) at the repository level.
\emph{First}, at the commit level, we explore the overlap between agent/\-non-agent test commits and mock/non-mock commits across all 169,361 test commits.
To assess whether these two categorical variables are statistically associated, we also applied a Chi-squared test of independence.
\emph{Second}, at the repository level, we group commits belonging to the same repository.
We categorize repositories based on their level of agentic activity:
(1) repositories with lower agentic activity (i.e.,~between 10 and 49 agent commits) and 
(2) repositories with higher agentic activity (i.e.,~at least 50 agent commits).
In both cases, we select repositories with at least 10 test commits to ensure a minimum level of testing activity.
We make no assumption about the presence of mocks in test commits, as this is the phenomenon we aim to investigate.
To assess the mock differences between agents and non-agents within the \emph{same} repository, we applied a paired Wilcoxon test and Cliff’s delta to compute the effect size.
\textbf{Rationale:} This analysis aims to better understand the extent to which coding agents contribute to mocking generation as compared to non-agents.
It is important to note that a mock overuse can make the test harder to understand and maintain, and they may provide less assurance that the code is working properly~\cite{google_mock}, but the ratios of mocking usage can vary across repositories.
This motivated our repository-level analysis, which performs a paired comparison of mock generation by agents and non-agents within the \emph{same} repository.

\subsubsection{RQ3. What types of mocks are generated by coding agents?}

In our final research question, we explore the presence of different
mock types, i.e.,~dummy, stub, spy, mock, and fake~\cite{meszaros2007xunit}, in mock commits.
Here, we consider repositories that contain at least one mock commit made by coding agents, totaling 496 repositories.
\textbf{Rationale:} We aim to explore the types of mocks created by agents and non-agents. This analysis helps us understand whether mock types are used evenly or tend to concentrate on specific types.

\section{Results}
\label{sec:results}

\subsection{RQ1. Coding Agents and Tests}

In this analysis, we explore all 1,254,878 commits of the 2,168 repositories.
Table~\ref{tab:contingency-test} presents the contingency table for the agent commits and test commits.
Overall, we detected 48,563 agent commits (in 1,219 repositories), of which 11,035 are test commits, resulting in a test commit ratio of 23\%.
By comparison, non-coding agents had a test commit ratio of only 13\% (i.e.,~158,326 out of 1,206,315).
Those 11,035 agent test commits are distributed across 729 repositories, indicating that 60\% (729 out of 1,219) of the repositories with agent activity also contain agent test activity.

\begin{table}[h]
    \centering
    \footnotesize
    \caption{Contingency table of agent commits versus test commits (all repositories).}
    \begin{tabular}{lccc}
        \toprule
        & \textbf{Test commit} & \textbf{Non-test commit} & \textbf{Total} \\
        \midrule
        \textbf{Agent commit}   & 11,035 & 37,528 & 48,563 \\
        \textbf{Non-agent commit} & 158,326 & 1,047,989 & 1,206,315 \\
        \midrule
        \textbf{Total}  & 169,361 & 1,085,517 & 1,254,878 \\
        \bottomrule
    \end{tabular}
    \label{tab:contingency-test}
\end{table}

\begin{figure}[h]
     \centering
         \includegraphics[width=0.29\textwidth]{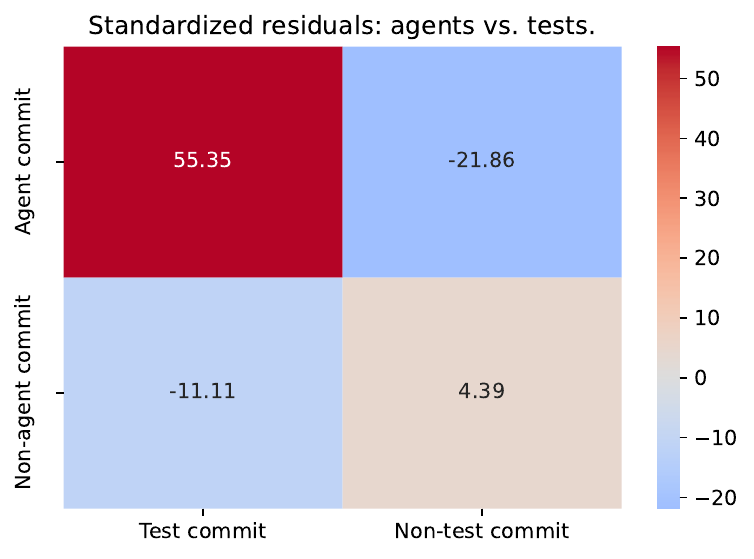}
         \caption{Standardized residuals for the contingency table of agent commits versus test commits (all repositories).}
         \Description{Standardized residuals for the contingency table of agent commits versus test commits (all repositories).}
        \label{fig:residuals-tests}
\end{figure}

We applied the Chi-squared test of independence to assess whether the two categorical variables are statistically associated.
The test revealed a significant association between them ($\chi^2$ = 3,683.06, df = 1, p < 0.001). The standardized residuals presented in Figure~\ref{fig:residuals-tests} show that the cell corresponding to agent commits and test commits is substantially overrepresented (55.35), whereas the remaining cells are under- or overrepresented to a lesser extent.
Overall, these results demonstrate a strong deviation from independence between the two variables, suggesting that agent commits are more likely to add or modify test files compared to non-agent commits.

\begin{center}
\fcolorbox{black}{gray!15}{
  \parbox{\dimexpr1\linewidth-2\fboxsep-2\fboxrule}{
\textbf{Observation 1}: 
60\% of the repositories with agent activity also contain agent test activity. 
23\% of commits made by coding agents involve the addition or modification of test files, compared with 13\% of commits made by non-agents.
  }
}
\end{center}


Table~\ref{tab:rq1-metrics} presents the ratio of test commits made by coding agents.
We observe a small variation across languages, with Python repositories showing a higher ratio (27\%) compared to JavaScript/TypeScript (21\%).
Among the coding agents, Copilot presents the highest proportion of test commits (27\%), followed by Claude (24\%) and Cursor (16\%).
It is worth noting, however, that Cursor accounts for the smallest number of agent commits (1,688) in our dataset.

\begin{table}[h]
    \centering
    \footnotesize
    \caption{Ratio of test commits made by coding agents.}
    \label{tab:rq1-metrics}
    \begin{tabular}{llccc}
        \toprule
        \multirow{2}{*}{\textbf{Category}} & \multirow{2}{*}{\textbf{Group}} & \multicolumn{3}{c}{\textbf{Agent}} \\
        \cmidrule(lr){3-5}
         &  & \textbf{Commits} & \textbf{Test Commits} & \textbf{Ratio (\%)} \\
        \midrule
        All & -- & 48,563 & 11,035 & 23\% \\
        \midrule
        \multirow{2}{*}{Language} 
            & Python  & 11,943 & 3,249 & 27\% \\
            & JS/TS   & 36,620 & 7,786 & 21\% \\  
        \midrule
        \multirow{4}{*}{Coding Agent} 
            & Copilot & 5,986 & 1,616 & 27\% \\
            & Claude  & 31,370 & 7,396 & 24\% \\
            & Cursor  & 1,688 & 278 & 16\% \\
            & Other & 9,746 & 1,851 & 19\% \\
        \bottomrule
    \end{tabular}
\end{table}


Interestingly, Table~\ref{tab:contingency-test} also shows that agent test commits account for 7\% of all test commits (i.e.,~11,035 out of 169,361).
For comparison, Table~\ref{tab:contingency-test-2025} presents the contingency table for repositories created in 2025.
In this case, we notice that agent test commits account for 17\% of all test commits (i.e.,~4,526 out of 26,654).
Thus, repositories created in 2025 present a higher proportion of agent test commits than the overall dataset (17\% vs. 7\%).

\begin{table}[h]
    \centering
    \footnotesize
    \caption{Contingency table of agent commits versus test commits (only repositories created in 2025).}
    \begin{tabular}{lccc}
        \toprule
        & \textbf{Test commit} & \textbf{Non-test commit} & \textbf{Total} \\
        \midrule
        \textbf{Agent commit}   & 4,526 & 18,654 & 23,180 \\
        \textbf{Non-agent commit} & 22,128 & 20,1744 & 22,3872 \\
        \midrule
        \textbf{Total}  & 26,654 & 220,398 & 247,052 \\
        \bottomrule
    \end{tabular}
    \label{tab:contingency-test-2025}
\end{table}

\begin{center}
\fcolorbox{black}{gray!15}{
  \parbox{\dimexpr1\linewidth-2\fboxsep-2\fboxrule}{
\textbf{Observation 2}: 
Considering all commits containing tests, 7\% are made by coding agents.
For repositories created in 2025, this proportion increases to 17\%.
  }
}
\end{center}


\subsection{RQ2. Coding Agents and Mocks}

To better understand how coding agents generate mocks, we conduct two complementary analyses: (1) at the commit level and (2) at the repository level.

\subsubsection{Commit-Level Analysis}

In this first analysis, we focus on the 169,361 test commits.
Table~\ref{tab:contingency-mock} presents the contingency table for the agent test commits and mock commits.
Overall, we detected 11,035 agent test commits (in 729 repositories), of which 3,934 are mock commits, resulting in a mock commit ratio of 36\%.
By comparison, non-coding agents presented a mock commit ratio of 26\% (i.e.,~40,966 out of 158,326).
Those 3,934 agent mock commits are distributed across 496 repositories, indicating that 68\% (496 out of 729) of the ones with agent test activity also contain agent mock activity.

\begin{table}[h]
    \centering
    \footnotesize
    \caption{Contingency table of agent test commits versus mock commits (all repositories).}
    \begin{tabular}{lccc}
        \toprule
        & \textbf{Mock commit} & \textbf{Non-mock commit} & \textbf{Total} \\
        \midrule
        \textbf{Agent test commit}   & 3,934 & 7,101 & 11,035 \\
        \textbf{Non-agent test commit} & 40,966 & 117,360 & 158,326 \\
        \midrule
        \textbf{Total}  & 44,900 & 124,461 & 169,361 \\
        \bottomrule
    \end{tabular}
    \label{tab:contingency-mock}
\end{table}

\begin{figure}[h]
     \centering
         \includegraphics[width=0.29\textwidth]{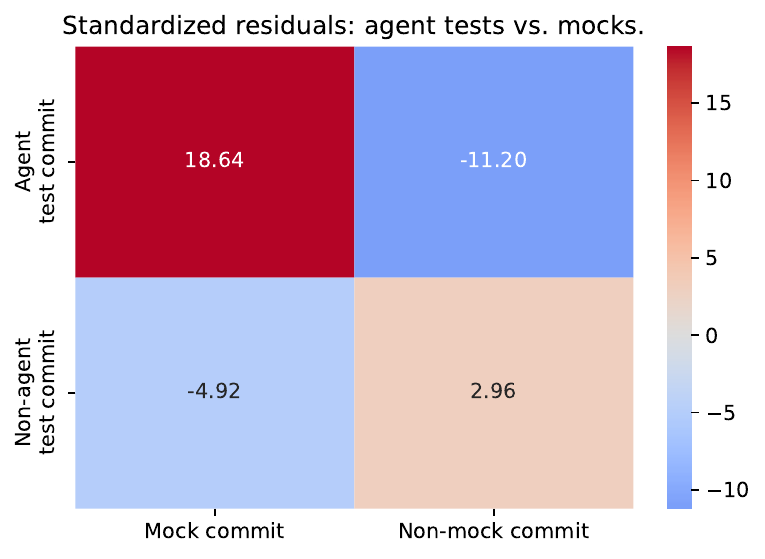}
         \caption{Standardized residuals for the contingency table of agent test commits versus mock commits (all repositories).}
         \Description{Standardized residuals for the contingency table of agent test commits versus mock commits (all repositories).}
        \label{fig:residuals-mocks}
\end{figure}

As in RQ1, we applied the Chi-squared test of independence.
In this case, the test also revealed a significant association ($\chi^2$ = 505.5, df = 1, p < 0.001).
The standardized residuals presented in Figure~\ref{fig:residuals-mocks} show that the cell corresponding to agent test commits and mock commits is substantially overrepresented (18.64).
Overall, these results suggest that agent test commits are more likely to add mocks compared to non-agent test commits.

\begin{center}
\fcolorbox{black}{gray!15}{
  \parbox{\dimexpr1\linewidth-2\fboxsep-2\fboxrule}{
\textbf{Observation 3}: 
Overall, 68\% of the repositories with agent test activity also contain agent mock activity.
36\% of test commits made by coding agents add mocks to test code, compared with 26\% of test commits made by non-agents.
  }
}
\end{center}


Table~\ref{tab:rq2-metrics} presents the ratio of mock commits made by coding agents in tests, separated by language and by agent.
We observe minimal differences between Python (37\%) and JavaScript/TypeScript (35\%) repositories.
Likewise, the ratios are similar across coding agents, with Cursor showing the highest proportion (37\%), followed by Copilot (36\%) and Claude (34\%).
It is worth noting, however, that Cursor contributes the smallest number of agent test commits (only 278) in our dataset.

\begin{table}[h]
    \centering
    \footnotesize
    \caption{Ratio of mock commits made by coding agents.}
    \label{tab:rq2-metrics}
    \begin{tabular}{llccc}
        \toprule
        \multirow{2}{*}{\textbf{Category}} & \multirow{2}{*}{\textbf{Group}} & \multicolumn{3}{c}{\textbf{Agent}} \\
        \cmidrule(lr){3-5}
         &  & \textbf{Test Commits} & \textbf{Mock Commits} & \textbf{Ratio (\%)} \\
        \midrule
        All & -- & 11,035 & 3,934 & 36\% \\
        \midrule
        \multirow{2}{*}{Language} 
            & Python  & 3,249 & 1,214 & 37\% \\
            & JS/TS   & 7,786 & 2,720 & 35\% \\
        \midrule
        \multirow{4}{*}{Coding Agent} 
            & Cursor  & 278 & 103 & 37\% \\
            & Copilot & 1,616 & 582 & 36\% \\
            & Claude  & 7,396 & 2,501 & 34\% \\
            & Other & 1,851 & 806 & 44\% \\
        \bottomrule
    \end{tabular}
\end{table}


\begin{center}
\fcolorbox{black}{gray!15}{
  \parbox{\dimexpr1\linewidth-2\fboxsep-2\fboxrule}{
\textbf{Observation 4}: 
Python, JavaScript, and TypeScript repositories present similar ratios of mock commits made by coding agents. Likewise, the ratios are comparable across different coding agents.
  }
}
\end{center}


In Table~\ref{tab:contingency-mock}, we notice that agent mock commits account for 9\% of all mock commits (i.e.,~3,934 out of 44,900).
For comparison, Table~\ref{tab:contingency-mock-2025} presents the contingency table for repositories created in 2025.
In this case, we notice that agent mock commits account for 19\% of all mock commits (i.e.,~1,529 out of 7,855).
Thus, repositories created in 2025 present a higher proportion of agent mock commits than the overall dataset (19\% vs. 9\%).

\begin{table}[h]
    \centering
    \footnotesize
    \caption{Contingency table of agent test commits versus mock commits (repositories created in 2025).}
    \begin{tabular}{lccc}
        \toprule
        & \textbf{Mock commit} & \textbf{Non-mock commit} & \textbf{Total} \\
        \midrule
        \textbf{Agent test commit}   & 1,529 & 2,997 & 4,526 \\
        \textbf{Non-agent test commit} & 6,326 & 15,802 & 22,128 \\
        \midrule
        \textbf{Total}  & 7,855 & 18,799 & 26,654 \\
        \bottomrule
    \end{tabular}
    \label{tab:contingency-mock-2025}
\end{table}

\begin{center}
\fcolorbox{black}{gray!15}{
  \parbox{\dimexpr1\linewidth-2\fboxsep-2\fboxrule}{
\textbf{Observation 5}: 
Considering all commits containing mocks, 9\% are made by coding agents.
For repositories created in 2025, this proportion increases to 19\%.
  }
}
\end{center}


\subsubsection{Repository-Level Analysis}

In this second analysis, we focus on the repository level by grouping commits belonging to the same repository.
We categorize repositories based on their level of agentic activity:
(1) repositories with lower agentic activity (i.e.,~between 10 and 49 agent commits) and 
(2) repositories with higher agentic activity (i.e.,~at least 50 agent commits).
In both cases, we only include repositories that contain at least 10 test commits to avoid those without minimal test activity.
Table~\ref{tab:rq2-repo} summarizes the median frequency of mock commits per repository for both categories.

\begin{table*}[h]
    \centering
    \scriptsize
    \caption{Median frequency of mock commits per repository.}
    \label{tab:rq2-repo}
    
    \begin{subtable}[t]{\textwidth}
        \centering
        \caption{Repositories with lower agentic activity (between 10 and 49 agent commits and at least 10 test commits).}
        \begin{tabular}{lrrrrrrrrr}
            \toprule
            \multirow{2}{*}{\textbf{Language}} 
            & \multirow{2}{*}{\textbf{\#Repositories}} 
            & \multicolumn{3}{c}{\textbf{Agent}} 
            & \multicolumn{3}{c}{\textbf{Non-Agent}} 
            & \multirow{2}{*}{\textbf{p-value}} 
            & \multirow{2}{*}{\textbf{Effect-Size}} \\ 
            \cmidrule(lr){3-5} \cmidrule(lr){6-8}
            &  & \textbf{Test Commits} & \textbf{Mock Commits} & \textbf{\%}
            & \textbf{Test Commits} & \textbf{Mock Commits} & \textbf{\%}
            &  &  \\
            \midrule
            JS/TS  & 195 & 4 & 1 & 25\% & 64 & 12 & 24\% & -- & -- \\
            Python & 87 & 8 & 2 & 27\% & 64 & 11 & 24\% & -- & -- \\
            \midrule
            All    & 282 & 5 & 1 & 27\% & 64 & 12 & 24\% & 0.002 & negligible \\
            \bottomrule
        \end{tabular}
        \label{tab:rq2-repo1}
    \end{subtable}
    
    \vspace{1em} 

    \begin{subtable}[t]{\textwidth}
        \centering
        \caption{Repositories with higher agentic activity (at least 50 agent commits and at least 10 test commits).}
        \begin{tabular}{lrrrrrrrrr}
            \toprule
            \multirow{2}{*}{\textbf{Language}} 
            & \multirow{2}{*}{\textbf{\#Repositories}} 
            & \multicolumn{3}{c}{\textbf{Agent}} 
            & \multicolumn{3}{c}{\textbf{Non-Agent}} 
            & \multirow{2}{*}{\textbf{p-value}} 
            & \multirow{2}{*}{\textbf{Effect-Size}} \\ 
            \cmidrule(lr){3-5} \cmidrule(lr){6-8}
            &  & \textbf{Test Commits} & \textbf{Mock Commits} & \textbf{\%}
            & \textbf{Test Commits} & \textbf{Mock Commits} & \textbf{\%}
            &  &  \\
            \midrule
            JS/TS  & 123 & 28 & 9 & 35\% & 42 & 7 & 22\% & -- & -- \\
            Python & 56 & 33 & 12 & 36\% & 105 & 24 & 30\% & -- & -- \\
            \midrule
            All    & 179 & 30 & 10 & 36\% & 56 & 13 & 28\% & <0.001 & small \\
            \bottomrule
        \end{tabular}
        \label{tab:rq2-repo2}
    \end{subtable}

\end{table*}

First, Table~\ref{tab:rq2-repo1} presents the analysis for the 282 repositories with lower agentic activity.
On the median, coding agents made 5 test commits per repository, of which 1 was a mock commit, resulting in a mock ratio of 27\%.
In contrast, non-agents produced a mock ratio of 24\%.
We applied a paired Wilcoxon test to compare the mock ratios between agents and non-agents within the same repository.\footnote{Normality tests indicated that the data were not normally distributed. Both Shapiro-Wilk and D’Agostino tests yielded p-values $<$ 0.001, thus, we used the non-parametric Wilcoxon test.}
The test revealed a significant difference with p-value = 0.0029.
However, the effect size was negligible (Cliff's delta = 0.002), indicating that the observed difference is statistically significant, but not meaningful.

Second, Table~\ref{tab:rq2-repo2} details the analysis for the 179 repositories with higher agentic activity.
In this case, on the median, coding agents made 30 test commits per repository, of which 10 were a mock commit, resulting in a mock ratio of 36\%.
In contrast, non-agents produced a mock ratio of 28\%.
Again, we applied a paired Wilcoxon test to compare the mock ratios between agents and non-agents.
The test revealed a significant difference with p-value $<$ 0.001 and, in this case, a small effect size (Cliff's delta = 0.252).

Table~\ref{tab:rq2-repo} also presents the median values per programming language.
We find no notable difference between Python and JavaScript/TypeScript.
For example, the mock commit ratios are 36\% for Python and 35\% for JavaScript/TypeScript in repositories with higher agentic activity.

\begin{center}
\fcolorbox{black}{gray!15}{
  \parbox{\dimexpr1\linewidth-2\fboxsep-2\fboxrule}{
\textbf{Observation 6}: 
In repositories with higher agentic activity, coding agents had a higher ratio of mock commits (36\%) compared to non-agents (28\%).
  }
}
\end{center}


\subsubsection{Examples of Repositories with High Frequency of Agent Mock Commits}

Table~\ref{tab:top10} presents repositories with a high frequency of agent mock commits.
Category \emph{All} in Table~\ref{tab:top10} presents the top-10 repositories with the most mock commits made by coding agents across the entire dataset.
The top-1 is promptfoo/promptfoo (148 agent mock commits), a tool for testing LLM applications.
As an example of an agent mock commit, we show commit 0cdc01a,\footnote{\url{https://github.com/promptfoo/promptfoo/commit/0cdc01aca0}} which added multiple mock identifiers to tests, such as \texttt{Mocked\-Function} and \texttt{createMock\-Target\-Provider}.
The second repository is drivly/ai (138), an agentic workflow platform.
As an example, we present commit 4eb5ea0,\footnote{\url{https://github.com/drivly/ai/commit/4eb5ea0415}} which states ``\emph{Add comprehensive unit tests for functions.do SDK}'' and includes diverse mock identifiers, as \texttt{mock\-\-Fetch}, \texttt{mockApi\-Instance}, and \texttt{consoleSpy}.
The third repository is liam-hq/liam (73), a tool to automatically generate ER diagrams from databases.
As an example, we show commit 695e79f,\footnote{\url{https://github.com/liam-hq/liam/commit/695e79f3eb}} which mentions ``\emph{Add comprehensive test coverage for all routing scenarios}'' and includes multiple occurrences of \texttt{createMockState}.

We also investigated the presence of agent mock commits in repositories maintained by top organizations on GitHub.
For this purpose, we collected the top 100 organizations with the most stars from Gitstar Ranking.\footnote{\url{https://gitstar-ranking.com/organizations}}
 Among them, we identified 24 repositories with at least one mock commit made by code agent: Microsoft (12), Home-Assistant (3), Redis (2), Azure (2), GitHub (1), Apache (1) MUI (1), Ethereum (1), and Cloudflare (1).
Table~\ref{tab:top10} presents the top-10 repositories with the most agent mock commits in the top organizations.
We notice important repositories such as home-assistant/core (61 agent mock commits) and microsoft/vscode (14).

In home-assistant/core, we present commit 4ca1ae6,\footnote{\url{https://github.com/home-assistant/core/commit/4ca1ae61aa}} which adds multiple mocks as \texttt{FAKE\_STATIONS} and \texttt{mocked\_stations}.
In microsoft/vscode, we show commit 8222326,\footnote{\url{https://github.com/microsoft/vscode/commit/822232673f178e733b0}} which includes mocks as \texttt{create\-MockService\-Config} and \texttt{mock\-Configuration\-True}.
Lastly, in apache/superset, we present commit 7d0a472,\footnote{\url{https://github.com/apache/superset/commit/7d0a472d1e401df6a92b}} which add mocks as \texttt{mockGet\-Bootstrap\-Data} and \texttt{mock\-Match\-Media}.

\begin{center}
\fcolorbox{black}{gray!15}{
  \parbox{\dimexpr1\linewidth-2\fboxsep-2\fboxrule}{
\textbf{Observation 7}: 
Coding agents made commits with mocks in a wide range of repositories, including 24 maintained by popular organizations such as Microsoft and Redis.
  }
}
\end{center}


\begin{table*}[h]
\centering
\scriptsize
\caption{Top-10 repositories with the most mock commits across the entire dataset and within top organizations.}
\begin{tabular}{l l l r r r}
\toprule
\multirow{2}{*}{\textbf{Category}} & \multirow{2}{*}{\textbf{Repository}} & \multirow{2}{*}{\textbf{Language}} & \multicolumn{3}{c}{\textbf{Agent}} \\
\cmidrule(lr){4-6}
 & & & \textbf{Test Commits} & \textbf{Mock Commits} & \textbf{Ratio} \\ \midrule
 \multirow{10}{*}{All} 
 & promptfoo/promptfoo & TypeScript & 232 & 148 & 0.64 \\
 & drivly/ai & TypeScript & 395 & 138 & 0.35 \\
 & liam-hq/liam & TypeScript & 511 & 73 & 0.14 \\
 & czlonkowski/n8n-mcp & TypeScript & 221 & 67 & 0.30 \\
 & chatman-media/timeline-studio & TypeScript & 144 & 66 & 0.46 \\
 & calcom/cal.com & TypeScript & 130 & 64 & 0.49 \\
 & home-assistant/core & Python & 87 & 61 & 0.70 \\
 & Kilo-Org/kilocode & TypeScript & 102 & 58 & 0.57 \\
 & dyoshikawa/rulesync & TypeScript & 189 & 57 & 0.30 \\
 & ar-io/ar-io-node & TypeScript & 106 & 45 & 0.42 \\
\midrule
\multirow{10}{*}{Top Organizations} 
& home-assistant/core & Python & 87 & 61 & 0.70 \\
& microsoft/fluidframework & TypeScript & 43 & 16 & 0.37 \\
& microsoft/vscode & TypeScript & 78 & 14 & 0.18 \\
& redis/agent-memory-server & Python & 55 & 12 & 0.22 \\
& microsoft/vscode-pull-request-github & TypeScript & 14 & 8 & 0.57 \\
& home-assistant/supervisor & Python & 11 & 7 & 0.64 \\
& cloudflare/workers-sdk & TypeScript & 19 & 7 & 0.37 \\
& apache/superset & TypeScript & 20 & 7 & 0.35 \\
& microsoft/omnichannel-chat-sdk & TypeScript & 15 & 7 & 0.47 \\
& Azure/LogicAppsUX & TypeScript & 11 & 5 & 0.45 \\
\bottomrule
\end{tabular}
\label{tab:top10}
\end{table*}

\subsection{RQ3. Types of Generated Mocks}

\begin{figure}[h]
     \centering
         \includegraphics[width=0.48\textwidth]{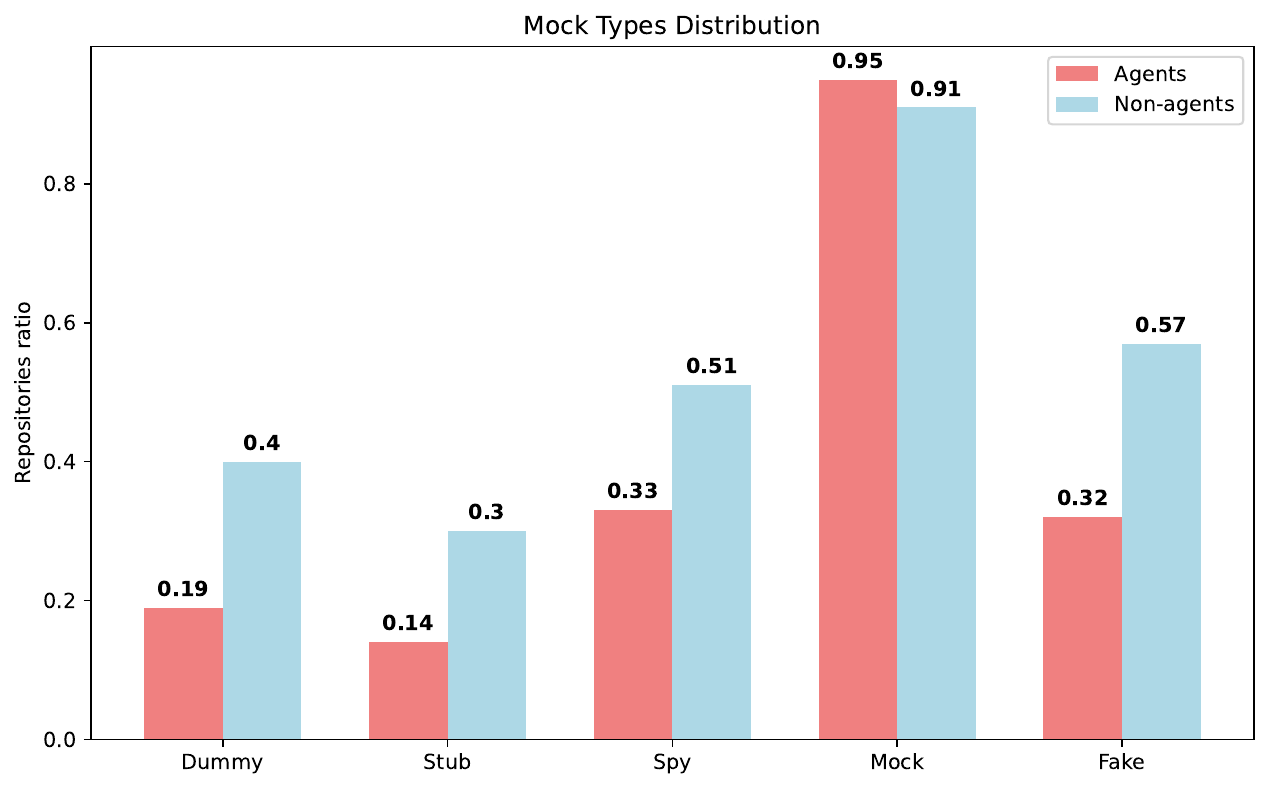}
         \caption{Distribution of mock types.}
         \Description{Distribution of mock types.}
        \label{fig:mock-types}
\end{figure}

In our final research question, we explore the presence of different mock types~\cite{meszaros2007xunit} (namely dummy, stub, spy, mock, and fake) within mock commits.
For this analysis, we consider only repositories that contain at least one mock commit made by coding agents, totaling 496 repositories.
Figure~\ref{fig:mock-types} shows the distribution of mock types across the 496 repositories.
For example, 19\% of the repositories contain at least one agent mock commit with a \emph{dummy}, compared to 40\% for non-agent commits.

The most commonly adopted type is \emph{mock} for both agents and non-agents, which seems to reflect the fact that developers tend to use the terms mocks and test doubles interchangeably~\cite{pereira2020assessing}.
Moreover, \emph{dummy} and \emph{stub} are the least adopted types, which are the simplest test doubles.
We recall that, in theory, a \emph{dummy} has no predefined or verified behavior, whereas a \emph{stub} only provides predefined implementations~\cite{fazzini2022use}.
It is worth noting that non-agent mock commits are more evenly distributed across types, particularly \emph{mock} (91\%), \emph{fake} (57\%), and \emph{spy} (51\%).
In contrast, agent mock commits are more concentrated in the \emph{mock} type (95\%), followed by \emph{spy} (33\%) and \emph{fake} (32\%).
This suggests that coding agents tend to generate less diverse test doubles as compared to non-agents.

\begin{center}
\fcolorbox{black}{gray!15}{
  \parbox{\dimexpr1\linewidth-2\fboxsep-2\fboxrule}{
\textbf{Observation 8}: 
Coding agents predominantly use the \emph{mock} type (95\%), whereas non-agents employ a wider variety: \emph{mock} (91\%), \emph{fake} (57\%), and \emph{spy} (51\%).
  }
}
\end{center}


\section{Discussion and Implications}
\label{sec:discussion}

\subsection{Coding agents are likely to add/modify tests}
We detected that agent commits are more likely to add or modify test files compared to non-agent commits (RQ1). 
Overall, 60\% of the repositories with agent activity also contain agent test activity, and 23\% of commits made by coding agents involve the addition or modification of test files, compared with 13\% of commits made by non-agents.
The Chi-squared test of independence showed that the condition agent commits and test commits is overrepresented.

\textbf{Implication:} The higher likelihood of coding agents modifying or adding test files suggests that these tools are being actively used not only for implementing production code but also for maintaining and expanding tests.
This indicates a growing potential for agents to support software testing tasks autonomously, which may influence how practitioners instruct coding agents to create tests and review the automatically generated tests.

\subsection{Coding agents are likely to add mocks}

We detected that agent test commits are more likely to add mocks to tests compared to non-agent test commits (RQ2). 
Overall, 68\% of the repositories with agent test activity also contain agent mock activity, and 36\% of test commits made by coding agents add mocks, compared with 26\% of test commits made by non-agents.
The Chi-squared test of independence showed that the condition agent test commits and mock commits is overrepresented.
Even within the same repository, we found that coding agents tend to add proportionally more mocks (36\%) than non-agents (28\%).
Using mocks in tests only guarantees success if the mocks match the real implementations.
In practice, this is hard to ensure, especially as the real code evolves and possibly gets out of sync with the mocks~\cite{google_mock}.

\textbf{Implication:}
Coding agents tend to rely more heavily on mocking when generating or modifying tests.
This may reflect an overuse of isolation techniques, potentially leading to tests that are easier to generate automatically, but less effective at validating real interactions~\cite{google_mock}. 
Practitioners should review such agent-generated tests to ensure appropriate use of mocks and adequate test realism.
This also opens opportunities for future research to assess whether such mocks impact test quality, maintainability, and coverage.

We recall that developers can customize the coding agent configuration files (e.g.,~\texttt{CLAUDE.md}), including commands, guidelines, and test instructions~\cite{claude-best-practices, copilot-best-practices, cursor-best-practices}.
These configuration files can also contain mock-related information, such as mock instructions, mocking best practices, mock anti-patterns, mock examples, and mock tools.
For example, the \texttt{CLAUDE.md} file of the repository browser-use/browser-use mentions: ``\emph{Never mock anything in tests, always use real objects!! The only exception is the LLM [...]}''.\footnote{\url{https://github.com/browser-use/browser-use/blob/0e925468bf/CLAUDE.md}}
We found only five agent mock commits in this repository; 3 are indeed related to LLMs,\footnote{\url{https://github.com/browser-use/browser-use/commit/cedb0d9cd4}} but 2 not seem to be in the scope of LLMs.\footnote{\url{https://github.com/browser-use/browser-use/commit/c16bfa4f65}}
Likewise, the \texttt{copilot-instructions.md} of home-assistant/core contains: ``\emph{Mock external APIs - Use fixtures with realistic JSON data}''.\footnote{\url{https://github.com/home-assistant/core/blob/fa86148df0/.github/copilot-instructions.md}}
Interestingly, this repository is among the ones with the most agent mock commits (61).
The file \texttt{AGENTS.md} of apache/superset is very specific on what should be used:``\emph{Mock patterns: Use MagicMock() for config objects, avoid AsyncMock for synchronous code}''.\footnote{\url{https://github.com/apache/superset/blob/98fba1eefe/AGENTS.md}}
Table~\ref{tab:coding-agent-files} details the frequency of test and mock occurrences in coding agent configuration files according to the GitHub Code Search.
It shows that the presence of test-related instructions is relatively common, whereas mock-related instructions are less frequent.

\begin{table}[h]
    \centering
    \footnotesize
    \caption{Frequency of test and mock occurrences in coding agent configuration files on GitHub (October 2025).}
    \begin{tabular}{lccc}
        \toprule
        \textbf{File} & \textbf{\#} & \textbf{With test} & \textbf{With mock} \\
        \midrule
        \texttt{CLAUDE.md} 
        & \href{https://github.com/search?q=path%3Aclaude.md&type=code}{\textcolor{blue}{112k}} 
        & \href{https://github.com/search?q=path%3Aclaude.md+test&type=code}{\textcolor{blue}{102k}} 
        & \href{https://github.com/search?q=path%3Aclaude.md+%28dummy+OR+stub+OR+spy+OR+mock+OR+fake%29&type=code&ref=advsearch}{\textcolor{blue}{13k}} \\
        \texttt{copilot-instructions.md} 
        & \href{https://github.com/search?q=path%3A.github%2Fcopilot-instructions.md}{\textcolor{blue}{44k}} 
        & \href{https://github.com/search?q=path%3A.github%2Fcopilot-instructions.md+test&type=code}{\textcolor{blue}{27k}} 
        & \href{https://github.com/search?q=path%3A.github%2Fcopilot-instructions.md+%28dummy+OR+stub+OR+spy+OR+mock+OR+fake%29&type=code}{\textcolor{blue}{7k}} \\
        \texttt{CURSOR.md} 
        & \href{https://github.com/search?q=path%3Acursor.md}{\textcolor{blue}{4.8k}} 
        & \href{https://github.com/search?q=path%3Acursor.md+test&type=code}{\textcolor{blue}{1.3k}} 
        & \href{https://github.com/search?q=path%3Acursor.md+%28dummy+OR+stub+OR+spy+OR+mock+OR+fake%29&type=code}{\textcolor{blue}{200}} \\
        \bottomrule
    \end{tabular}
    \label{tab:coding-agent-files}
\end{table}

\textbf{Implication:}
Given the tendency of coding agents to rely on mocking, practitioners should include guidance on mocking best practices and anti-patterns in agent configuration files to ensure more consistent test generation.
Future research can leverage coding agent configuration files to explore whether the provided instructions are actually reflected in the agent-generated code.

\subsection{Recent repositories contain a higher proportion of agent test and mock commits}

In RQs 1 and 2, we also analyzed the proportion of test and mock commits in repositories created more recently, in 2025.
We found that repositories created in 2025 present a higher proportion of agent test commits than the overall dataset (17\% vs. 7\%).
Likewise, repositories created in 2025 present a higher proportion of agent mock commits than the overall dataset (19\% vs. 9\%).
Given these trends, it is plausible that in the near future, a substantial portion of tests (and mocks) will be generated by coding agents.

\textbf{Implication:}
The increasing share of test and mock commits by coding agents in recently created repositories suggests that agent-generated testing is becoming more common.
This opens room for future research to investigate how the increasing presence of coding agents affects the evolution of testing practices over time.

\subsection{Coding agents use less diverse test doubles}

The testing literature proposes five test doubles: dummy, stub, spy, mock, and fake, each one with its characteristics~\cite{meszaros2007xunit}.
In RQ3, we found that agents predominantly use the \emph{mock} type (95\%), whereas non-agents employ a wider variety: \emph{mock} (91\%), \emph{fake} (57\%), and \emph{spy} (51\%).
In both cases, \emph{dummy} and \emph{stub} are the least adopted types.

\textbf{Implication:}
The narrow use of test double types by coding agents may suggest limited diversity in their testing strategies.
Practitioners should be aware that agent-generated tests may rely on a single form of test double (i.e.,~mock), potentially missing opportunities to simulate different isolation scenarios.
Practitioners can provide more fine-grained instructions in agent configuration files about the rationales behind the usage of each test double.

\section{Threats to Validity}

\noindent\textbf{Detection of test and mock commits:}
To detect test commits, we analyzed multiple test file and directory patterns, which were extracted from the official documentation of popular testing frameworks, such as Pytest, unittest, Jest, Sinon.JS, and Jasmine.
Our solution to detect mock commits was inspired by prior studies~\cite{pereira2020assessing, fazzini2022use}, which also rely on code identifiers to detect the presence of test doubles in test code.
Thus, based on the fact that we rely on official documentation and established solutions to detect tests and mocks, we minimize the risks of false positive test and mock commits.

\noindent\textbf{Detection of agent commits:}
We started the process to detect agent commits by selecting repositories that contained agent-related files or directories associated with Claude, Copilot, or Cursor, which are three leading agents~\cite{li2025rise}.
These three agents work by authoring or co-authoring commits.
Next, we analyzed commit authors and co-authors to identify the presence of multiple coding agents, as described in Section~\ref{sec:detecting-agents}.
This step was necessary because repositories may employ multiple coding agents (e.g.,~Claude and Aider), each leaving distinct traces in their commits.
Therefore, we minimize the risks of false negatives, that is, not finding real agent commits.

\noindent\textbf{Agent commits with co-authors:}
Developers can attribute a commit to more than one author by adding \texttt{Co-authored-by:} trailers to the commit message~\cite{co-authored-by}.
Although GitHub recommends using the \texttt{Co-authored-by:}~\cite{co-authored-by}, we observed that coding agents may also use variants such as \texttt{Co-Authored-By:}, as reported in Section~\ref{sec:detecting-agents}.
Thus, we searched for these trailers in commit messages in a case-insensitive manner to reduce the risks of false negatives.
Moreover, when a commit includes multiple co-authors, it becomes less clear how individual contributions are distributed.
To minimize the risks, we reported multiple views of the data in Tables~\ref{tab:rq1-metrics} and~\ref{tab:rq2-metrics}.

\noindent\textbf{Generalization of the results}. 
We analyzed more than 1.2 million commits made in 2025 across 2,168 repositories written in three programming languages (Python, JavaScript, and TypeScript).
Moreover, we mined commits to identify the usage of numerous coding agent tools, including Claude, Copilot, Cursor, Aider, OpenHands, and Devin AI.
Despite these observations, our findings -- as usual in empirical software engineering -- cannot be directly generalized to repositories written in other languages or using other agents.

\section{Related Work}

\subsection{Coding Agents}
Since coding agents are very new~\cite{agentminingpaper}, few empirical studies of them exist (benchmark evaluations of coding agents, or studies of code completion LLMs are out of scope for this paper). Cursor, a coding agent integrated in a variant of Visual Studio Code \cite{cursor}, has been both popular and available for longer than other leading agents. It has been the focus of two studies. A controlled experiment by Becker et al.~\cite{becker2025measuring} focused on estimated and actual task completion time. Sixteen developers worked on 246 tasks in several large projects. The developers estimated that they were 20\% more productive using Cursor. Surprisingly, the experiment found that task completion time \emph{increased} by 19\%, contrary to developer perception. The second study is an observation study by Kumar et al.~\cite{kumar2025sharp}, where 19 developers also used Cursor, working on 33 issues. They observed various collaboration strategies between developers and agents, from complete ``one-shot'' delegation, where developers send the entire task to the agent, up to more collaborative processes, where the task is divided into simpler subtasks that the agent works on.

Agent interaction logs are explored by Bouzenia and Pradel~\cite{bouzenia2025understanding}. These logs come from less established tools (OpenHands~\cite{wang2024openhands}, Autocoderover~\cite{zhang2024autocoderover}, and RepairAgent~\cite{bouzenia2024repairagent}) and are extracted from the SWE-bench benchmark, so they do not reflect real usage. The study analyzed 120 interaction logs submitted to the benchmark, comparing logs from successful and unsuccessful tasks across a variety of factors, such as the number of interaction steps and tokens used, the type of actions they took, and anti-patterns in failing logs.

\subsection{Software Tests and Test Doubles}

Ideally, test suites should have good quality to catch more bugs and protect against regressions~\cite{aniche2021developers, hora2024test, hora2025exceptional, dalton2020exceptional, marcilio2021java}.
When creating tests, developers may find dependencies that make the test harder to implement~\cite{pereira2020assessing}.
In this scenario, they can use test doubles (also known as mocks) to emulate the dependencies' behavior, contributing to making the test fast, isolated, and deterministic~\cite{meszaros2007xunit, pereira2020assessing, spadini2017mock, spadini2019mock, fazzini2022use}.
There are five types of test doubles: dummy, stub, spy, mock, and fake~\cite{meszaros2007xunit}.
There exists a vast literature exploring mocking practices mainly in Java~\cite{mostafa2014empirical, spadini2017mock, spadini2019mock, pereira2020assessing, li2024automatically}, but also in Android~\cite{fazzini2022use} and Python~\cite{trautsch2017there}.
In the context of LLMs and mocking, an early study evaluated OpenAI's GPT-4o for automating mock decisions in Apache Dubbo.
In a controlled experiment, the study compared model outputs with developer choices, finding that the model tends to generate more mocks than developers~\cite{qin2025mock}.
In our research, we leverage the traces left by agents in real-world software systems to study mock generation in 2,168 repositories.

\section{Conclusion}

This paper presented an empirical study to investigate the presence of mocks in agent-generated tests of real-world software systems.
We analyzed over 1.2 million commits made in 2025 in 2,168 TypeScript, JavaScript, and Python repositories.
Overall, we find that coding agents are more likely to modify tests and to add mocks to tests than non-coding agents.
Finally, we concluded by discussing multiple implications for researchers and practitioners, such as the need to review agent-generated
tests to ensure appropriate use of mocks and the call for guidance on mocking best practices and anti-patterns in agent configuration files.

As future work, we plan to conduct a more qualitative investigation of agent-generated tests and mocks, for example, to assess their quality and type, verify whether some mocks could be avoided, and identify whether they introduce harmful testing practices.
Finally, we plan to explore the content of coding agent files (e.g.,~\texttt{CLAUDE.md}) to better understand their role in software testing and mocking.

\begin{acks}
This research was supported by CNPq, CAPES, FAPEMIG, and the French State.
This work was partially supported by INES.IA (National Institute of Science and Technology for Software Engineering Based on and for Artificial Intelligence), www.ines.org.br, CNPq grants: 408817/2024-0 and 403304/2025-3.
This study also received financial support from the French State in the framework of the Investments for the Future programme IdEx université de Bordeaux.
\end{acks}

\bibliographystyle{ACM-Reference-Format}
\bibliography{main}

\end{document}